\documentclass[%
 aip,
 amsmath,amssymb,
 reprint,%
]{revtex4-1}

\usepackage{graphicx}% Include figure files
\usepackage{dcolumn}% Align table columns on decimal point
\usepackage{bm}% bold math
\usepackage[utf8]{inputenc}
\usepackage[T1]{fontenc}
\usepackage{mathptmx}
\usepackage{etoolbox}

\makeatletter
\def\@email#1#2{%
 \endgroup
 \patchcmd{\titleblock@produce}
  {\frontmatter@RRAPformat}
  {\frontmatter@RRAPformat{\produce@RRAP{*#1\href{mailto:#2}{#2}}}\frontmatter@RRAPformat}
  {}{}
}%
\makeatother
\begin{document}

\preprint{AIP/123-QED}

\title{An integrated ultrahigh vacuum cluster tool for diamond surface science and single nitrogen-vacancy center measurements}
% Force line breaks with \\
\author{Zhiyang Yuan}
\author{Sorawis Sangtawesin}
\author{Lila V. H. Rodgers}
% \altaffiliation[Now at ]{MIT Lincoln Lab.}
\author{Kalliope Zervas}
\author{James J. Allred}
\author{Jared Rovny}
\affiliation{Department of Electrical and Computer Engineering, Princeton University, Princeton, New Jersey 08544, USA}

\author{Patryk Gumann}
\affiliation{IBM T.J. Watson Research Center, Yorktown Heights, New York 10598, USA}

\author{Nathalie P. de Leon}
\email{npdeleon@princeton.edu.}
\affiliation{Department of Electrical and Computer Engineering, Princeton University, Princeton, New Jersey 08544, USA}

\date{\today}% It is always \today, today,

\begin{abstract}
% Your abstract should not contain displayed equations, footnotes, references, graphics, or tables. It should be one paragraph of 250 words providing a summary of the new information, results of general interest, and conclusions
% 05/19/2026: 114 words
We present a custom-designed ultrahigh vacuum (UHV) cluster tool developed for studying shallow nitrogen–vacancy (NV) centers in diamond, enabling in situ diamond surface preparation, characterization, and single NV center dynamics measurements within a single connected platform. The system combines a surface science chamber for controlled surface modification and analysis with a cryogenic confocal microscope chamber dedicated to NV spin and optical measurements. This integrated approach enables a direct correlation between diamond surface chemistry and the resulting NV spin and charge properties. The instrument provides a versatile platform for systematic studies of surface-induced decoherence mechanisms and charge dynamics for shallow NV centers, and establishes a pathway toward reproducible surface engineering for quantum sensing applications.
\end{abstract}

\maketitle

\section{\label{introduction}Introduction}

Nitrogen–vacancy (NV) centers in diamond have emerged as a leading solid-state platform for quantum sensing and quantum information applications due to their long spin coherence times, optical addressability, and compatibility with ambient and cryogenic operation.\cite{Lovchinsky2016, Maurer2012, Hensen2015, schirhagl2014nitrogen} 
In particular, shallow NV centers---located within a few nanometers of the diamond surface---enable high-sensitivity nanoscale sensing of magnetic,\cite{taylor2008high,Sage2013, rovny2024nanoscale} electric,\cite{dolde2011electric, iwasaki2017direct} and thermal signals\cite{kucsko2013nanometre, neumann2013high} from external systems. However, the diamond surface can host contaminants and defects that shorten shallow NV lifetimes and cause NV charge-state instability.\cite{Sangtawesin2019a,Bluvstein2019, yuan2020charge, Myers2014} 

Previous works have demonstrated that diamond surface chemistry and structure play a critical role in determining the properties of shallow NV centers. Surface termination, adsorbates, reconstruction, and subsurface damage can all affect NV center properties.\cite{janitz2022diamond,Sangtawesin2019a, ristein2006surface, barry2020sensitivity, zuo2023impact, nagura2026understanding} As a result, reproducible control and detailed characterization of the diamond surface are essential for understanding and mitigating surface-related decoherence mechanisms. Most NV experiments rely on ex situ surface treatments followed by optical measurements performed in separate systems, with unavoidable exposure to ambient conditions between preparation and characterization steps. The presence of uncontrolled surface adsorbates complicates the direct correlation between surface properties and NV center behavior, and leaves unaddressed the question of how much surface noise can be attributable to extrinsic surface adsorbates versus intrinsic surface defects.

To address this challenge, we have designed, constructed, and validated an ultrahigh vacuum (UHV) cluster tool that integrates diamond surface modification and characterization with cryogenic confocal microscopy for NV measurements. The system consists of a surface science chamber dedicated to diamond surface preparation and analysis and a confocal microscope chamber that supports optical access and spin-resonance experiments on NV centers. Crucially, the UHV environment supports a low adventitious carbon contamination rate, with pristine surfaces remaining contamination-free for more than one month. This architecture allows diamond samples to be prepared, characterized, and measured without breaking vacuum, preserving surface conditions that are otherwise difficult to maintain across separate experimental platforms, enabling direct correlation between diamond surface properties and shallow NV performance. Furthermore, this integrated platform provides a general framework for investigating other color centers and defects in diamond, as well as defects in a broad range of solid-state host materials, by enabling controlled surface engineering and in-situ characterization of defect properties under well-defined environments.

In this paper, we describe the design, implementation, and capabilities of this integrated UHV cluster tool. We describe the vacuum chamber architecture, sample handling, the cryogenic confocal microscope, and surface science instrumentation. Particular emphasis is placed on design choices that enable compatibility among UHV conditions, cryogenic and room temperature operation, high temperature surface preparation, high Rabi frequency microwave driving, and high numerical aperture (NA) confocal optical access. We further demonstrate the performance of the system through representative measurements on diamond surfaces and shallow NV centers, illustrating its utility for studying surface-induced effects relevant to quantum information science applications.

\section{\label{instrument_design}Instrument Design}

\subsection{\label{instru:overview}Overview}

\begin{figure*}
    \centering
    \includegraphics{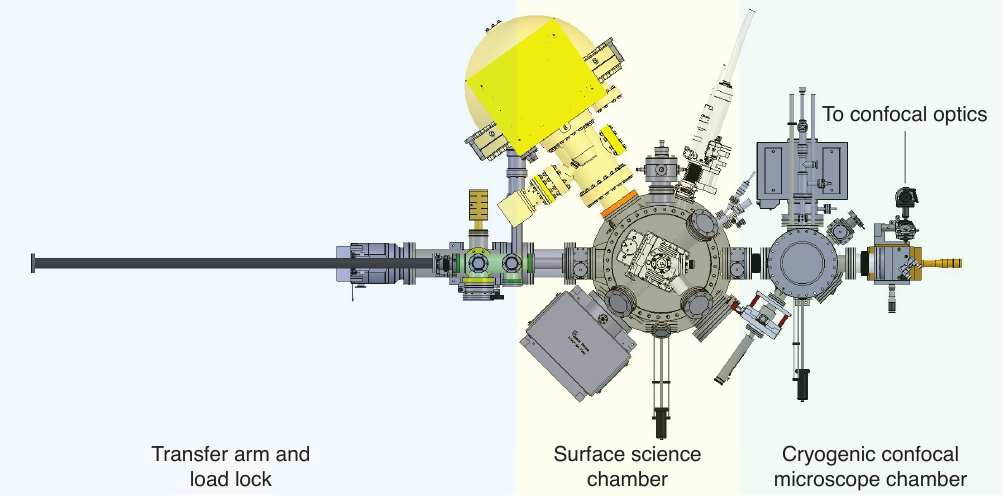}
    \caption{Schematic (top view) of the ultrahigh vacuum (UHV) cluster tool for diamond surface science and nitrogen-vacancy (NV) center characterization. The system is configured in a linear cluster geometry and comprises three main modules: a load lock, a surface science chamber, and a cryogenic confocal microscope chamber. To accommodate multiple tools, the surface science chamber has a body outer diameter (OD) of 16.5~in. (419~mm) and a height of 18~in. (457~mm). The specific tools are labeled and described in Fig.~\ref{fig:main_chamber}. The cryogenic confocal microscope chamber has a body OD of 10~in. (254~mm) and a height of 17.5~in. (445~mm). }
    \label{fig:chambers}
\end{figure*}

The UHV cluster tool comprises three interconnected chambers (Fig.~\ref{fig:chambers}): a load-lock chamber for introducing samples from ambient conditions, a surface science chamber dedicated to diamond surface modification and characterization, and a cryogenic confocal microscope chamber for NV measurements.
The surface science chamber is equipped with X-ray photoelectron spectroscopy (XPS) for chemical state analysis, low-energy electron diffraction (LEED) for surface crystallography, a heating stage for UHV annealing, and a gas cracker for introducing reactive atomic species to terminate the surface.\cite{dontschuk2023x, Sangtawesin2019a, zulkharnay2024oxygen} The cryogenic confocal microscope chamber enables detailed characterization of NV centers and their local noise environment under UHV conditions across a temperature range from liquid helium temperature to room temperature. The setup supports measurements of spin dynamics including spin–lattice relaxation ($T_1$), spin dephasing ($T_2$), and coherence under dynamical decoupling sequences for noise spectral decomposition, providing quantitative insight into surface-related noise mechanisms.\cite{Sangtawesin2019a,myers2017double,Romach2015, Myers2014} The surface science chamber and the cryogenic confocal microscope chamber are maintained at approximately $5\times10^{-10}$~mbar, ensuring the diamond surface remains clean throughout the experiments. Residual gas analyzer (RGA) measurements indicate that the base pressure is dominated by hydrogen, while the partial pressures of water and hydrocarbons are more than two orders of magnitude lower (see Appendix \ref{vacuum_details} for vacuum details).

The three chambers are arranged in a linear cluster configuration and mounted on an optical table together with the optical components of the confocal microscope (Fig.~\ref{fig:optical_setup}).
Diamond samples are mounted on custom-machined holders attached to standardized flag plates compatible with all chambers in the tool (Fig.~\ref{fig:cryo_chamber_top}(c)). Sample transfer between chambers is performed using a magnetically coupled transfer arm equipped with two sample slots, enabling efficient handling of multiple samples without breaking vacuum. Within each chamber, the sample holder can be manipulated using a wobble stick designed to interface with the flag plate geometry, allowing reliable pickup and placement on chamber-specific stages. This unified mounting and transfer scheme ensures reproducible sample positioning and seamless operation across all modules.

A typical experiment proceeds by loading a diamond sample through the load lock, performing surface preparation and characterization in the surface science chamber, and transferring the sample under UHV to the cryogenic confocal microscope for NV measurements, enabling direct correlation between surface conditions and NV properties. In the following subsections, we first describe the design of the cryogenic confocal microscope chamber and then the surface science chamber.

% Cryo chamber
% Top lid size: DN 200 CF, 10'' OD, 420GSG200 Pfeiffer
% Height: 17.5''

% Surface chamber
% Top flange: Lesker F1650X1400R, 16.5'' OD, DN350CF
% height 18''

\subsection{\label{instru:confocal_chamber}Cryogenic confocal microscope chamber}

\begin{figure}
    \centering
    \includegraphics{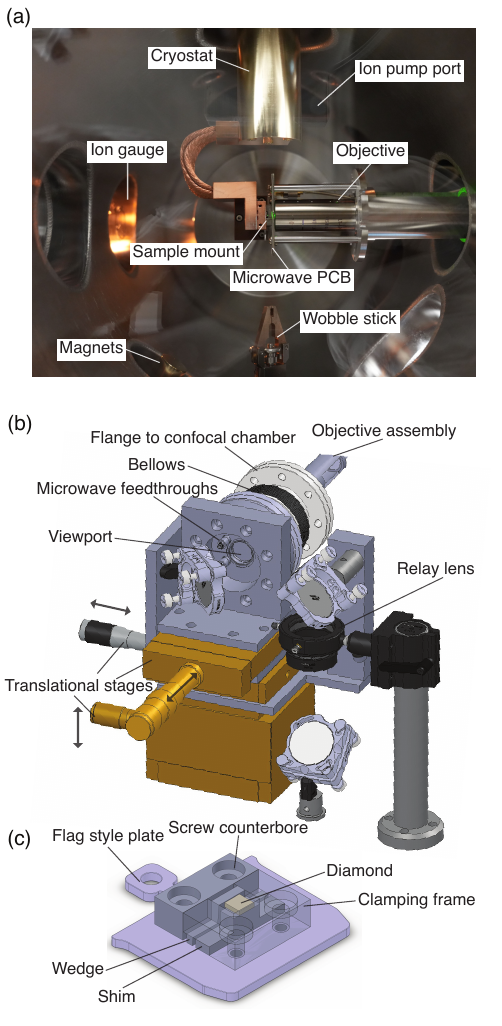}
    \caption{(a) Top view of the cryogenic confocal microscope chamber, imaged through the top viewport. (b) Design of the translation stages for the UHV confocal microscope. The objective assembly is mounted on the viewport flange, which is connected to the cryogenic confocal microscope chamber via a bellows to allow flexible movement. Two mirrors are mounted on different levels of the translation stages to preserve the optical alignment during stage movement. (c) Diamond sample mount for sample processing and confocal measurement. The clamping frame is rendered semi-transparent to show the underlying shim and wedge. Assembly parts are made of molybdenum. The assembly is mounted on a standard flag style plate through four tantalum screws. Tantalum is selected to prevent high-temperature diffusion bonding (galling) with the molybdenum components during thermal processing. The diamond shown in this figure measures $2 \times 2 \times 0.5$~mm$^3$. Different diamond sizes can be accommodated by replacing the wedge and shim with appropriately sized components.}
    \label{fig:cryo_chamber_top}
\end{figure}

The cryogenic confocal microscope chamber (Fig.~\ref{fig:cryo_chamber_top}) is designed to enable optical and spin measurements of single NV centers under UHV conditions. The chamber is connected to the surface science chamber via a gate valve, allowing diamond samples to be transferred under vacuum using the shared transfer arm. This configuration preserves surface cleanliness and termination during transfer from surface preparation and characterization to optical measurement.

There are three central design challenges in performing single NV center measurements in UHV conditions: high-NA imaging, microwave delivery for high Rabi frequency spin driving, and precise magnetic field alignment to the NV center quantization axis. The chamber is optimized for simultaneous high-NA imaging and high-Rabi-frequency microwave driving by imaging through an aperture in a microwave printed circuit board (PCB) with an objective that can be heated to $150\,^\circ\text{C}$. In-vacuum permanent magnets attached to an external positioning assembly allow for precise static field alignment. These elements allow full optical initialization, manipulation, and readout of NV spin states. The chamber is also coupled to a cryostat that provides stable low-temperature operation, enabling studies of temperature-dependent NV optical and spin properties.

\subsubsection{\label{instru:confocal_optics}Confocal optics}

\begin{figure*}
    \centering
    \includegraphics{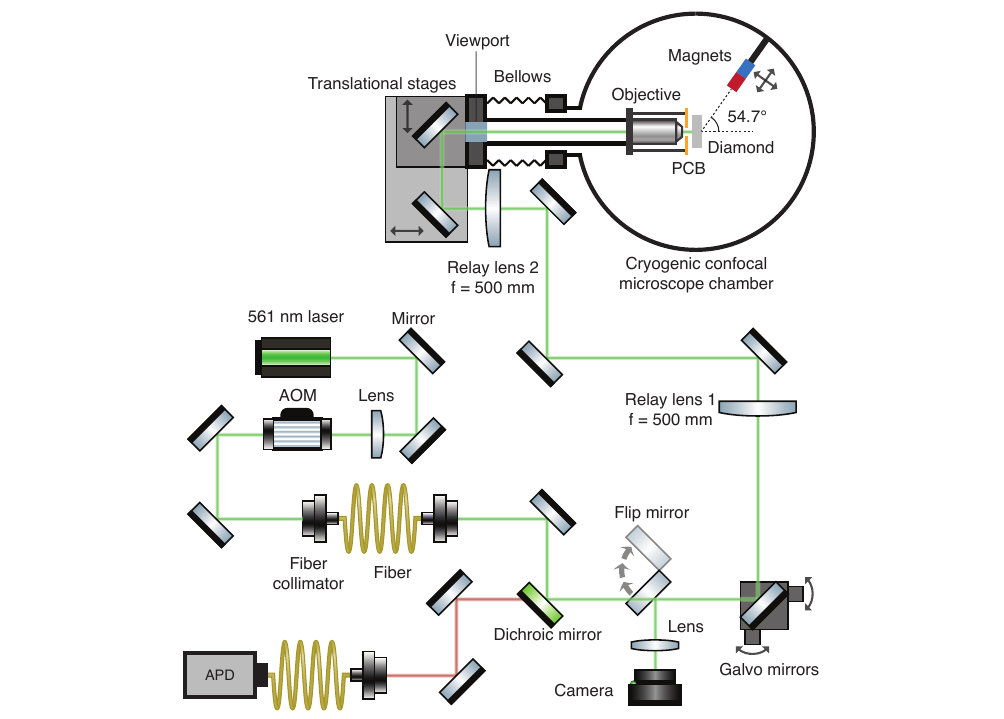}
    \caption{Optical setup of the confocal microscope. A 561~nm laser is coupled to an acousto-optic modulator (AOM) to enable pulsed optical excitation for fast spin readout of NV centers. The beam is focused onto the AOM with a plano-convex lens to achieve a rise time of around 20~ns. In the confocal configuration, a pair of galvanometer (galvo) mirrors scans the lateral beam position at the objective focal plane. Two relay lenses arranged in a 4-$f$~configuration image the scanning point from the galvo mirrors onto the back aperture of the UHV objective. Distances in the schematic are not to scale. The UHV objective can be translated relative to the diamond sample using external translation stages. The objective is coupled to the stages through a lens tube and a viewport flange, and mirrors are also mounted on the translation stages to maintain beam alignment during motion. NV fluorescence is collected by the same objective and propagates backward along the excitation path. A long-pass dichroic mirror separates the fluorescence from the excitation beam, and the signal is coupled into a single-mode fiber connected to an avalanche photodiode (APD). The single-mode fiber also serves as the confocal pinhole to reject out-of-focus background. Reflected green light can be directed to a camera to assist in focusing on the diamond surface; this imaging path is disabled during measurements using a flip mirror. The vertical optical path (see Fig.~\ref{fig:cryo_chamber_top}(b)) is simplified in this two-dimensional schematic for clarity.}
    \label{fig:optical_setup}
\end{figure*}

\begin{figure*}
    \centering
    \includegraphics{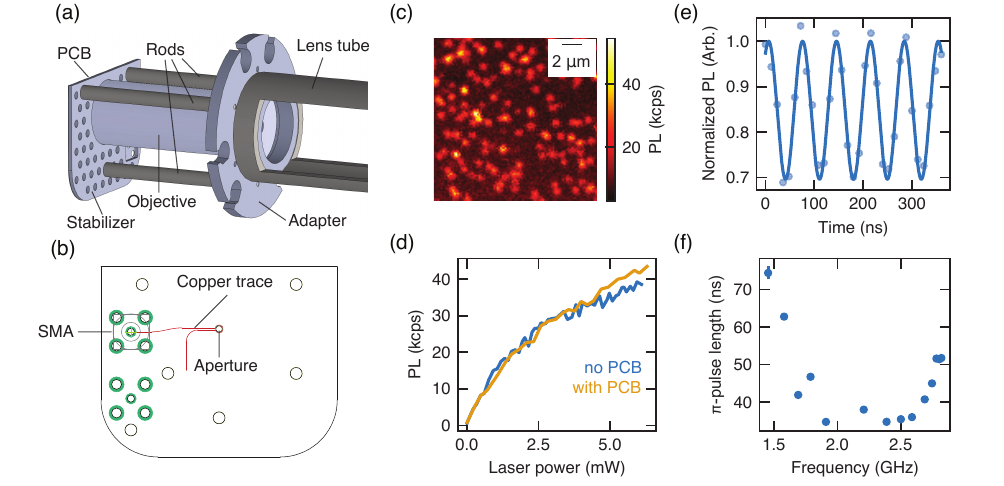}
    \caption{(a) PCB mount assembly. The PCB is mounted to a metallic plate for mechanical stability, which is connected to the adapter via three support rods. The plate has an opening larger than the objective diameter, enabling the PCB to be positioned in close proximity to the objective. Venting holes in the plate are designed to prevent gas trapping between the PCB and the plate under UHV conditions. The support rods are symmetrically arranged around the objective on the adapter. Openings in the adapter provide access for microwave cables. An aperture in the lens tube allows viewing of the back aperture of the objective during optical alignment. All components are fabricated from 316L stainless steel.
    (b) UHV microwave PCB design. ROGERS RO3010 laminate is used as the PCB substrate material. The copper trace has a width of $96.5\,\mu\text{m}$ and a thickness of $35\,\mu\text{m}$, corresponding to a characteristic impedance of 50~$\Omega$. The backside of the PCB is fully coated with $35\,\mu\text{m}$ thick copper to serve as a ground plane. The central aperture has a diameter of 1.2~mm. Additional openings accommodate screws for attaching the support rods and securing the PCB to the metal plate.
    (c) UHV confocal scan showing individual NV centers as bright spots. The excitation laser wavelength is 561~nm with a power of 1.3~mW. The acquisition time for each pixel is 5~ms. The total scan time is around 50~s.
    (d) NV photoluminescence (PL) saturation curves measured with (yellow) and without (blue) the PCB, indicating no significant obstruction of the excitation laser or emitted NV PL. 
    (e) NV Rabi oscillations obtained using the UHV PCB, yielding a Rabi $\pi$-pulse length of 35~ns at a microwave power of approximately 16~W. The microwave frequency is 2.388~GHz, resonant with the NV spin $m_s=0$ to $m_s=-1$ transition. Experimental data (points) are fitted to a sinusoidal function (solid line).
    (f) Frequency dependence of the Rabi $\pi$-pulse length, demonstrating efficient microwave driving of the NV spin from 1.5~GHz to 2.8~GHz. The applied microwave power is held constant at around 16~W for all measurements.}
    \label{fig:PCB_confocal}
\end{figure*}

High-NA optical access is essential for efficient excitation and collection of fluorescence from single NV centers.\cite{gruber1997scanning} To achieve the required NA under UHV conditions, the microscope objective is placed inside the cryogenic confocal microscope chamber (Fig.~\ref{fig:cryo_chamber_top}(a) and Fig.~\ref{fig:optical_setup}). The system employs an Attocube LT-APO/VISIR (NA = 0.82) objective designed for UHV compatibility and low-temperature operation. This objective provides apochromatic correction over the 565--770~nm range, closely matching the NV fluorescence spectrum. To optimize transmission and minimize chromatic aberration between the excitation and emission paths, a 561~nm laser is used instead of the more commonly employed 532~nm excitation. In addition, the collimation of the excitation beam is adjusted to compensate for the residual chromatic focal shift of the objective. The objective is also compatible with UHV bakeout, withstanding temperatures up to $150\,^\circ\text{C}$.
Optical excitation of the NV centers and collection of their fluorescence are performed through a UHV-compatible viewport equipped with an anti-reflection coating (550--1100~nm) optimized for the excitation and emission wavelengths. The viewport assembly also incorporates SMA electrical feedthroughs for microwave signals.

Conventional confocal microscopes implemented in high vacuum  environments often rely on multi-axis piezoelectric stages inside the vacuum chamber for raster scanning.\cite{schaefer2014diamond} Such approaches increase mechanical complexity within the UHV volume and complicate thermal management, particularly in the context of sample mounts that are compatible with high temperature annealing for surface preparation. In contrast, the present design adopts an alternative architecture based on external scanning optics. The objective assembly is mechanically coupled to a three-axis translation stage located outside the UHV chamber (Fig.~\ref{fig:cryo_chamber_top}(b) and Fig.~\ref{fig:optical_setup}). The optical assembly and translation stages are connected to the chamber via a flexible bellows, allowing controlled relative motion of the objective with respect to the sample while maintaining vacuum integrity.

The external translation stages provide coarse manual alignment of the objective relative to the diamond surface. Fine lateral scanning across the sample is achieved using a pair of galvo mirrors (Thorlabs, GVS012) located outside the vacuum chamber. The galvo and two relay lenses are arranged in a 4-$f$~optical configuration with respect to the objective back aperture, enabling diffraction-limited lateral scanning with minimal beam walk or aberrations at the sample plane. Precise axial scanning is implemented using a piezoelectric actuator (Thorlabs, PK4GA3P2) mounted on the translation stage. Because the objective assembly is mechanically coupled to the chamber through the bellows (Fig.~\ref{fig:optical_setup}), the axial translation stage and piezo are designed to operate against the vacuum-induced force acting on the viewport flange (DN63CF), which is estimated to be around 320~N for a 4.5~in. (114~mm) outer diameter (OD) flange with a bore diameter of 2.5~in. (64~mm).

With this configuration, diffraction-limited confocal imaging is achieved over a field of view of \(100 \times 100~\mu\mathrm{m}^2\). Different regions of the sample can be accessed by translating the objective using translation stages. The saturation PL count rate of a single NV center is measured to be 46.5~kcps, with a saturation power of 2.5~mW under the 561~nm laser excitation.

\subsubsection{\label{instru:sample_mount}Sample mount}
The sample mount (Fig.~\ref{fig:cryo_chamber_top}(c)) is designed to be compatible with both high-temperature surface processing in the surface science chamber and low-temperature optical measurements in the cryogenic confocal microscope chamber. These dual requirements impose significant constraints on material choice, geometry, and thermal integration. The mount must be UHV compatible, capable of withstanding annealing temperatures exceeding $1000\,^\circ\text{C}$, efficiently thermally coupled to the cryostat for cooling to liquid-helium temperatures, and compatible with the short working distance of the objective.

Molybdenum (Mo) is selected as the material for the sample mount due to its low vapor pressure, excellent UHV compatibility, high melting point ($2623\,^\circ\text{C}$), and mechanical stability at elevated temperatures. Its good thermal conductivity also facilitates efficient heat transfer during both annealing and cryogenic operation.

The diamond sample is secured by a side-clamping geometry, leaving the top surface fully exposed for surface processing, optical measurement, and microwave delivery. This configuration avoids the use of top clamps or adhesives that could interfere with the high-NA optical access. The geometry is further constrained by the short working distance (0.65~mm) of the UHV objective and the presence of the microwave PCB positioned between the objective and the sample. To prevent mechanical interference, the mount is designed such that no portion of the holder protrudes above the sample surface or contacts the PCB during operation.

To accommodate diamond substrates of varying thicknesses, interchangeable shims are placed beneath the sample. These shims are machined to different heights so that, regardless of sample thickness, the diamond surface remains the highest and most forward-facing element of the assembly. 

\subsubsection{\label{instru:PCB}Microwave PCB}

Microwave delivery for coherent control of NV spin states is implemented using a custom-designed UHV-compatible PCB integrated into the cryogenic confocal microscope chamber. Efficient microwave driving of NV center spin states requires the microwave antenna to be positioned in close proximity to the diamond surface while maintaining compatibility with UHV, cryogenic operation, and high-NA optical access.

To meet these requirements, the PCB is mounted as part of an assembly attached to the objective holder, allowing precise and reproducible positioning relative to the sample (Fig.~\ref{fig:PCB_confocal}(a)). During operation, the PCB is positioned between the objective and the diamond sample. By tuning the length of the support rods, the PCB is placed approximately 0.17~mm from the diamond surface, ensuring a finite gap to avoid mechanical contact while providing sufficiently strong microwave fields for efficient NV spin manipulation.

The PCB is implemented as a microstrip structure with a loop geometry surrounding a central aperture (Fig.~\ref{fig:PCB_confocal}(b)). This aperture allows transmission of the excitation laser and collection of NV fluorescence, preserving compatibility with high-NA confocal imaging. To accommodate the limited working distance of the objective (0.65~mm) and minimize optical obstruction, the PCB is fabricated using the thinnest practical substrate (thickness: 0.0050~$\pm$~0.0005~in. (0.13~$\pm$~0.01~mm)). To enhance mechanical rigidity and minimize bending during operation, a metallic plate is mounted behind the PCB. Mechanical stability of the assembly is verified through operation at high microwave powers. A similar PCB configuration has been described earlier in reference\cite{yuan2024instructional}.

A confocal scan of a diamond sample is shown in Fig.~\ref{fig:PCB_confocal}(c), where single NV centers appear as diffraction-limited spots. NV photoluminescence (PL) saturation curves are measured for two comparable NV centers with and without the PCB in place (Fig.~\ref{fig:PCB_confocal}(d)). The nearly identical saturation behavior confirms that the central aperture enables efficient transmission of both the excitation laser and the emitted NV PL.
The copper trace on the PCB is designed for efficient microwave transmission. The microwave signal reflects at the open end of the trace, forming standing waves along its length. The trace length is chosen such that antinodes occur near the central aperture at frequencies commonly used in NV experiments. As shown in Fig.~\ref{fig:PCB_confocal}(e,f), this design enables efficient driving of the NV spin, achieving $\pi$-pulse durations of 35--75~ns (corresponding to Rabi frequencies of 6.7--14.3~MHz and driving field amplitudes ($B_1$) of 2.4--5.1~G) over a frequency range of 1.5--2.8~GHz, in agreement with the numerical simulations \cite{yuan2024instructional}. In this configuration, the PCB is attached to the objective and remains centered relative to the field of view. It maintains a consistently high Rabi frequency driving when scanning across the sample, in contrast to other microwave delivery approaches such as fabricated transmission lines or wire bonds on the diamond, where the relative distance between the NV center and the antenna limits the area over which efficient microwave driving can be achieved.

\begin{figure}
    \centering
    \includegraphics{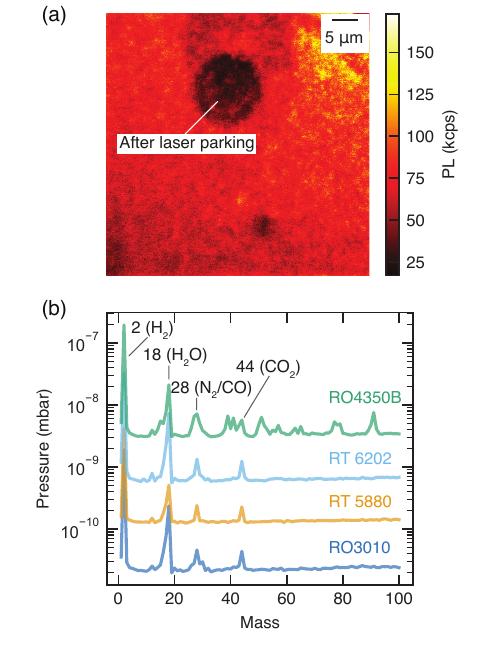}
    \caption{(a) A confocal scan of the diamond surface after high power microwave through a PCB made of RO4350B. The dark regions show the increased background surface PL can be quenched by scanning the green laser. (b) Averaged residual gas analyzer (RGA) mass spectra during continuous microwave operation for 2~hours showing outgassing partial pressures from different PCB substrate materials: blue: RO3010, yellow: RT/duroid 5880, light blue: RT/duroid 6202, green: RO4350B. The top three curves have been vertically shifted for clarity.}
    \label{fig:PCB_RGA}
\end{figure}

A critical design choice is the PCB material. During earlier chamber design iterations, increased chamber pressure and changes in diamond surface PL were observed during high-power microwave operation, indicating significant outgassing from the PCB substrate when heated by microwave dissipation (Fig.~\ref{fig:PCB_RGA}(a)). To mitigate this effect, four PCB substrate materials with low outgassing ratings, based on NASA outgassing data,\cite{campbell1975outgassing} were evaluated: ROGERS Corporation RO4350B, RT/duroid 6202, RT/duroid 5880, RO3010.

Outgassing tests were conducted in a dedicated UHV test chamber equipped with an RGA. For each PCB substrate material, the chamber was baked under identical conditions (around $130\,^\circ\text{C}$ for 48~hours) to ensure consistent starting conditions. Microwave signals were subsequently applied through UHV-compatible feedthroughs, and the resulting outgassing spectra were recorded using the RGA.
The averaged RGA mass spectra for the four PCB substrate materials are shown in Fig.~\ref{fig:PCB_RGA}(b). The substrate used in an earlier design, RO4350B (widely used in trapped ion experiments \cite{brown2007loading, guise2014vacuum, tanaka2009design}) exhibited significant outgassing at masses around 40 and above 50, likely associated with hydrocarbon species. In contrast, the other three substrate materials showed dominant peaks only at mass = 2 (H$_2$), 18 (H$_2$O), 28 (N$_2$ / CO), and 44 (CO$_2$). These species are not expected to induce measurable PL from the diamond surface under our experimental conditions.

Among the three PCB substrate materials that exhibit low outgassing, RO3010 has the lowest coefficient of thermal expansion, which helps minimize PCB deformation during microwave operation. Therefore, RO3010 was selected as the substrate material for use in the cryogenic confocal microscope chamber. The PCB demonstrated good mechanical stability, and no measurable diamond surface PL was observed during microwave operation.

\subsubsection{\label{instru:magnets}Magnetic field alignment}

\begin{figure}
    \centering
    \includegraphics{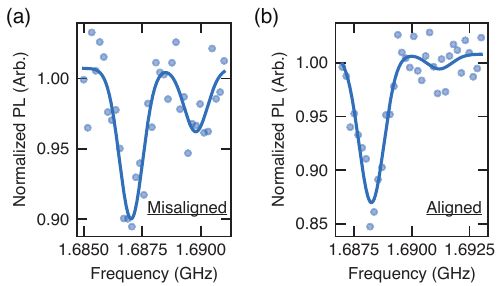}
    \caption{Pulsed optically detected magnetic resonance (ODMR) measurements of the $^{15}$N hyperfine splitting of the NV center when the magnetic field is misaligned (a) and aligned (b) with the NV axis. The magnets are moved laterally by 0.2~mm between (a) and (b). Microwave $\pi$-pulse width of 504~ns is used for these measurements. The magnetic field strength is 422~G.}
    \label{fig:hyperfine}
\end{figure}

A static bias magnetic field is applied in the cryogenic confocal microscope chamber using permanent magnets to enable NV center spin manipulations. Permanent magnets provide a compact UHV-compatible solution for generating stable magnetic fields without introducing electrical noise or additional heat load to the cryogenic environment.

Custom NdFeB disc magnets are used, with dimensions of 0.625~in. (15.88~mm) in diameter. The magnets are grade N45SH and are gold plated (Ni–Cu–Ni–Au) for improved corrosion resistance and UHV compatibility. This magnet grade has a maximum operating temperature of $150\,^\circ\text{C}$ and a Curie temperature of $340\,^\circ\text{C}$, allowing bakeout at temperatures required for UHV operation without degradation of magnetic performance.

The permanent magnets are mounted on a three-axis positioning assembly to allow precise control of the magnetic field magnitude and direction at the sample position. The XYZ manipulator consists of a Pfeiffer XY-axis precision manipulator (420MXY040-12) combined with a UHV Design linear bellows drive (LBD35-100-h) for axial positioning. This configuration enables reproducible positioning of the magnet assembly relative to the diamond sample while maintaining UHV integrity.

Within the cryogenic confocal microscope chamber, the magnet assembly is oriented at the magic angle ($54.7^\circ$) relative to the sample surface (Fig.~\ref{fig:cryo_chamber_top}(a) and Fig.~\ref{fig:optical_setup}). For (100)-oriented diamond samples, this geometry enables efficient alignment of the magnetic field with one of the NV axes. Using this setup, the magnetic field strength at the sample position can be continuously tuned from approximately 18~G to 525~G by adjusting the magnet–sample separation.

Magnetic field alignment and magnitude are verified using NV center spectroscopy. Alignment can be confirmed by monitoring the transition frequencies of NV centers with different orientations. Fine alignment near the excited-state level anticrossing (ESLAC) can be assessed through the polarization of the hyperfine splitting of the nitrogen nuclear spin states. Using these methods, an alignment precision better than $0.5^\circ$ relative to the NV axis can be achieved.\cite{jacques2009dynamic, bucher2019quantum} Example data showing the $^{15}$NV hyperfine-splitting polarizations for aligned and slightly misaligned magnetic fields are presented in Fig.~\ref{fig:hyperfine}. This approach provides a reliable and in situ method for calibrating the magnetic field for NV spin measurements.

\subsubsection{\label{instru:cryostat}UHV-compatible cryostat}
The cryogenic confocal microscope chamber is equipped with a UHV-compatible Janis ST-400 cryostat, enabling low-temperature measurements of NV centers under UHV conditions. The cryostat supports cooling of the sample to liquid-helium temperatures, providing access to temperature regimes relevant for studies of NV optical linewidths,\cite{fu2009observation} spin relaxation,\cite{cambria2023temperature} and thermally activated surface noise processes.

The cryostat cold head is connected to the sample mount via flexible copper braids (Fig.~\ref{fig:cryo_chamber_top}(a)), which provide efficient thermal conduction while minimizing the transmission of mechanical vibrations to the sample. This decoupling is essential for maintaining stable confocal alignment and reproducible optical measurements at low temperature.

Unlike conventional cryostats that initiate cooling under rough vacuum conditions, the UHV-compatible ST-400 allows the cooling process to be carried out while maintaining UHV in the measurement chamber. This capability enables cryogenic NV measurements to be performed on atomically clean and well-defined diamond surfaces prepared in situ, without exposure to ambient gases during cooldown.

With continuous helium flow, a base temperature of approximately 4~K is achieved at the cryostat cold head. High-resolution photoluminescence excitation (PLE) measurements of the NV zero-phonon line confirm that the diamond sample temperature remains below 20~K during operation. \cite{chu2014coherent, fu2009observation} The sample temperature was estimated from the PLE linewidth of a shallow implanted NV center (30~keV implantation energy). Because surface effects can contribute to additional linewidth broadening, this estimate represents a conservative upper bound, and the actual sample temperature is likely lower than 20~K. This combination of UHV compatibility and cryogenic operation provides a unique platform for studying the interplay between diamond surface chemistry, temperature, and NV center optical and spin properties.

\subsection{\label{instru:surface_science_chamber}Surface science chamber}

\begin{figure}
    \centering
    \includegraphics{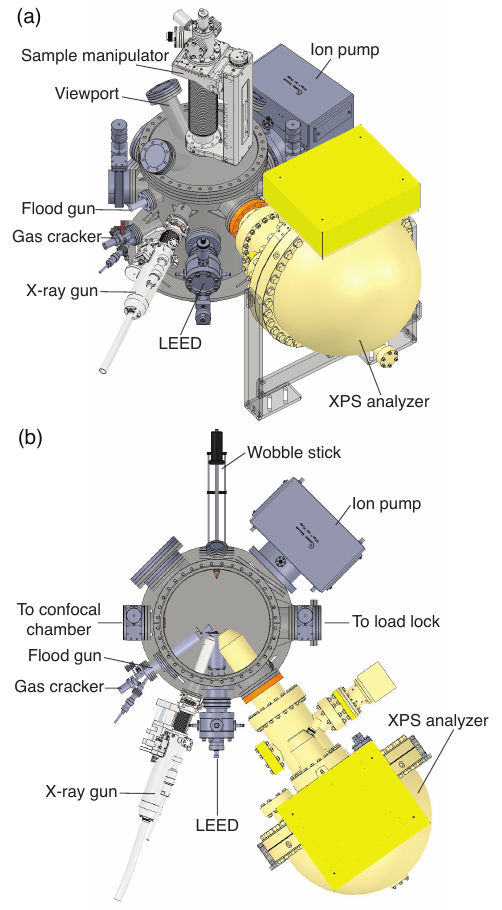}
    \caption{Schematic of the surface science chamber: (a) side view; (b) top view with the lid and sample manipulator removed. Surface science tools are mounted to flange ports welded into the chamber body. Several viewports are incorporated to facilitate sample manipulation and inspection within the chamber. The load-lock chamber and the cryogenic confocal microscope chamber are connected through gate valves and are not shown here.}
    \label{fig:main_chamber}
\end{figure}

The surface science chamber is designed for controlled modification and characterization of diamond surfaces under UHV conditions. Surface preparation capabilities include in situ heating, enabling removal of surface contamination and chemical desorption of diamond surfaces. A thermal gas cracker is integrated for controlled exposure to reactive species, allowing modification of the diamond surface termination.\cite{zulkharnay2024oxygen} Surface characterization is performed in situ using XPS and LEED, which provide complementary information on surface chemical composition and crystalline order, respectively. 

These tools are integrated into the surface science chamber through placement at different angular orientations and heights relative to the sample (see Fig. \ref{fig:main_chamber}). The diamond sample is mounted on a manipulator (UHV Design, MC-17315-001) that provides translational motion within the chamber as well as rotational control, enabling alignment with each characterization and processing tool. 

\subsubsection{\label{instru:heater}Heater}

A UHV-compatible heating system (HEAT3, PREVAC) is integrated into the sample manipulator to enable controlled thermal processing of diamond samples. The heater is mounted behind the sample plate on a dual-level sample receiver of the manipulator, providing direct radiative coupling to the diamond sample while maintaining compatibility with surface characterization tools. The heater foil assembly itself is mounted on a standardized flag plate, allowing it to be transferred, unloaded, and serviced through the sample transfer arm without venting the chamber.

The heating system supports two operating modes: resistive heating and electron-beam (e-beam) heating. In the resistive mode, the heater provides sufficient radiative power to raise the sample temperature to approximately $800\,^\circ\text{C}$. For higher-temperature processing, the heater can be operated in e-beam mode, enabling sample temperatures exceeding $1000\,^\circ\text{C}$. This dual-mode capability allows a wide range of thermal treatments to be performed within the same chamber.

The system temperature is monitored using a K-type thermocouple in thermal contact with the sample plate. Calibration between the thermocouple reading and the actual sample temperature was performed for both heating modes. Custom control software was developed to regulate the sample temperature ramp rates and dwell times during annealing, enabling highly repeatable thermal processing protocols. Moderate-temperature annealing above $300\,^\circ\text{C}$ is used to remove surface adsorbates from diamond samples, while higher-temperature annealing can be employed to desorb surface termination species and modify the surface chemical state.\cite{thomas1992thermal, su1998thermal} 

Thermal processing can be performed concurrently with in situ surface characterization using XPS or LEED, allowing direct observation of temperature-dependent surface processes. Representative XPS measurements illustrating oxygen desorption from the diamond surface during annealing are presented in Sec.~\ref{sec:oxygen_desorption}. Resistive heating is compatible with simultaneous XPS and LEED measurements; however, e-beam heating is disabled during XPS operation, as electrons emitted from the heater can saturate the XPS electron analyzer.

\subsubsection{\label{instru:gas_cracker}Thermal gas cracker}
The surface science chamber is equipped with a Mantis MGC75 Thermal Gas Cracker, designed to generate atomic species such as hydrogen and oxygen from molecular gas precursors under UHV conditions. The source operates via thermally activated ``catalytic'' dissociation: the precursor gas passes through a heated capillary, where collisions with the hot surfaces lead to a high degree of dissociation into reactive atomic species. This approach enables controlled delivery of atomic hydrogen, oxygen, or other species to the diamond surface for modification of surface terminations.

The thermal gas cracker provides a new avenue for diamond surface termination, complementing conventional methods such as gas-flow furnaces or plasma etching. Unlike high-temperature furnace treatments, the gas cracker allows in situ surface modification under UHV conditions, and unlike plasma etching, it generates reactive species without introducing significant subsurface damage. This capability is particularly valuable for preparing surfaces for shallow NV centers, where maintaining crystalline integrity and minimizing near-surface defects is critical.

\subsubsection{\label{instru:XPS}XPS}
XPS is implemented in the surface science chamber to enable in situ analysis of the chemical composition and termination of diamond surfaces. The XPS system consists of three primary components: an X-ray source, an electron flood gun, and a hemispherical electron energy analyzer.

The X-ray source is an RS 40B1 (PREVAC), equipped with a Mg/Al twin anode and dual filaments. This configuration enables selection between Mg K$\alpha$ (1253.4~eV) and Al K$\alpha$ (1486.6~eV) excitation and supports continuous operation at source powers of up to 300~W (Mg) and 400~W (Al), providing sufficient photon flux for surface-sensitive measurements under UHV conditions. Having two photon sources can help cross-check XPS spectrum peak assignments and identify Auger peaks. For radiation safety, the chamber viewports are fitted with lead-glass shielding. 

Because diamond is an insulating material and is susceptible to significant surface charging under X-ray illumination, a flood gun (FS40A, PREVAC) is incorporated to provide charge compensation during XPS measurements. The flood gun supplies low-energy electrons ($< 500\,\mathrm{eV}$) to neutralize accumulated positive charge on the sample surface, enabling reliable acquisition of core-level spectra without excessive peak shifting or broadening.

Photoelectrons emitted from the sample are analyzed using a SIGMA Surface Science Aspect electron analyzer, which provides energy-resolved detection of emitted electrons with high sensitivity. Together, these components enable quantitative analysis of the elemental composition and chemical bonding states at the diamond surface.

The three XPS components are mounted at the same vertical height within the surface science chamber to ensure proper alignment at the sample position. The X-ray source is installed at the magic angle of $54.7^\circ$ relative to the electron analyzer. During XPS operation, the diamond surface is oriented to face the electron analyzer to maximize signal collection and measurement sensitivity.

\begin{figure}
    \centering
    \includegraphics{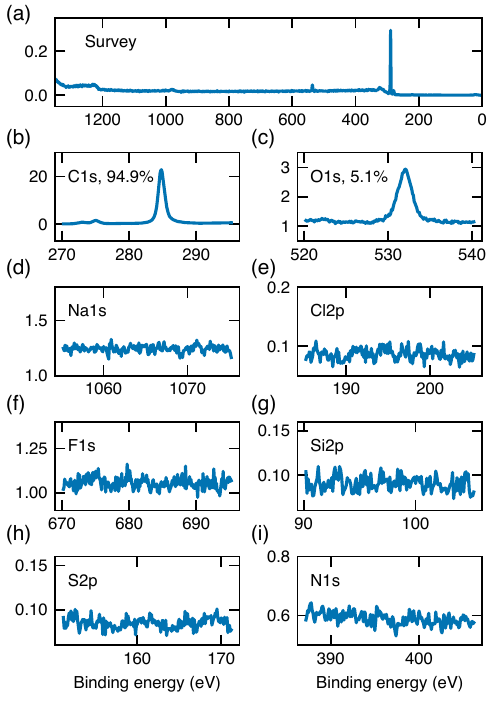}
    \caption{(a) X-ray photoelectron spectroscopy (XPS) survey scan and (b–i) high-resolution scans of individual core-level peaks for an oxygen-terminated diamond surface measured under UHV conditions. The spectra indicate a clean diamond surface with no detectable contamination. Atomic percentages of carbon and oxygen are extracted from the integrated peak areas.\cite{Sangtawesin2019a} Intensities in the survey and high-resolution scans are not on the same scale. XPS survey and high-resolution scans are acquired with analyzer dwell times of 0.2~s and 0.1~s, respectively.}
    \label{fig:XPS_scans}
\end{figure}

Representative XPS scans of a clean oxygen terminated diamond surface acquired by this tool are shown in Fig.~\ref{fig:XPS_scans}. XPS measurements are used to assess surface cleanliness, detect residual contamination, and verify surface termination following thermal annealing or exposure to atomic species generated by the thermal gas cracker. This capability is essential for correlating diamond surface chemistry with the optical and spin properties of shallow NV centers measured in the cryogenic confocal microscope chamber.

\subsubsection{\label{instru:LEED}LEED}

\begin{figure}
    \centering
    \includegraphics{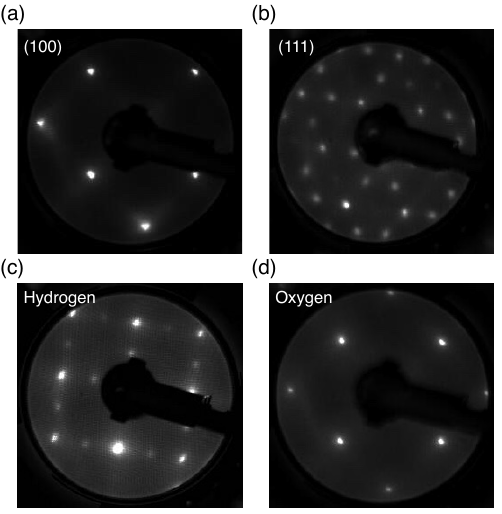}
    \caption{(a,b) Low-energy electron diffraction (LEED) patterns of (a) the diamond (100) surface at a beam energy of 100~eV and (b) the diamond (111) surface at 200~eV. The characteristic four-fold (square) symmetry in (a) and six-fold (hexagonal) symmetry in (b) clearly distinguish the two crystallographic orientations.
    (c,d) LEED patterns of hydrogen-terminated (100) diamond surface at 137~eV (c) and oxygen-terminated (100) diamond surface at 150~eV (d). The hydrogen-terminated surface exhibits the characteristic $2\times1$ reconstruction. The hydrogen and oxygen terminations are prepared by annealing the diamond samples in gas-flow furnaces.\cite{Sangtawesin2019a, zhang2023neutral, rodgers2024diamond}}
    \label{fig:LEED}
\end{figure}

The surface science chamber is also equipped with a LEED system (LPS075-D, OCI Vacuum Microengineering) for characterization of diamond surface structure and crystallographic order. The beam energy of this tool is tunable from 5 to 750~eV.

LEED provides direct information on surface periodicity, reconstruction, and long-range order, making it a complementary technique to XPS for diamond surface studies. In particular, LEED is used to distinguish between different diamond surface orientations, such as (111) and (100) (see Fig. \ref{fig:LEED} (a,b)), and to verify surface reconstruction following thermal annealing or surface termination procedures (see Fig. \ref{fig:LEED} (c,d)). The presence and symmetry of diffraction patterns allow assessment of surface quality and ordering prior to transfer to the cryogenic confocal microscope chamber for NV measurements.

\section{\label{results}Results}

\subsection{\label{sec:laser_surface_pl}Laser induced diamond surface PL}
\begin{figure}
    \centering
    \includegraphics{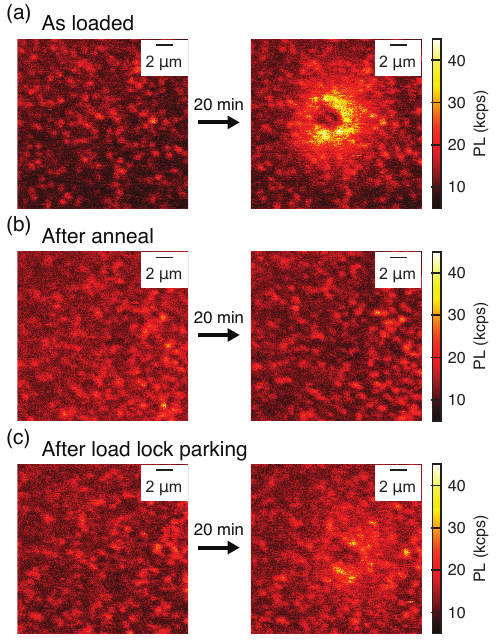}
    \caption{Confocal PL maps showing the evolution of the surface background under 561~nm excitation (1.45~mW). (a) As loaded: Continuous laser exposure for 20~minutes on an NV center creates a pronounced local background PL ``halo.'' (b) After UHV Anneal: No obvious increased surface background PL is observed after 1~hour annealing at $350\,^\circ\text{C}$ in the surface science chamber, indicating removal of surface adsorbates. (c) Load lock Exposure: After 19~hours in the high vacuum load lock, the increased surface background PL partially returns, suggesting re-contamination of the diamond surface even under high vacuum environment. In all cases, the excitation laser is parked on the NV center located at the center of each confocal map.}
    \label{fig:parking_PL}
\end{figure}

For diamond samples directly loaded into the cryogenic confocal microscope chamber, an increase in surface background PL intensity was observed in the region surrounding the laser parking position (Fig.~\ref{fig:parking_PL}(a)). Notably, this localized PL increase occurred only when the laser was parked on an NV center; no comparable effect was observed when parking the laser on positions without NV emission. After the formation of the increased surface background PL region, long laser exposure of that region could locally suppress the elevated surface background PL at the laser spot.

Following in situ annealing at $350\,^\circ\text{C}$ for 1~hour in the surface science chamber, we repeated the laser parking for another 20~min in a different area of the sample surface, and this surface background PL enhancement around the parking NV center spot was no longer observed (Fig.~\ref{fig:parking_PL}(b)). The suppression of the effect after moderate-temperature annealing suggests that the phenomenon is associated with surface adsorbates or contamination that can be removed by thermal treatment under UHV conditions.

Interestingly, the PL enhancement could be partially restored after re-exposing the sample to non-UHV environments. Following the UHV annealing in the surface science chamber, the sample was kept in the load-lock chamber for 19~hours. The load-lock chamber pressure remained around $6\times10^{-8}$~mbar during this period. Subsequent confocal measurements showed similar increased surface background PL near the parking position, although the magnitude of the increased PL was reduced compared to that of the as-loaded surface (Fig.~\ref{fig:parking_PL}(c)). 

The laser-induced surface PL enhancement varies for different diamond surfaces. In our measurements, a sample with an as-grown $^{12}$C-enriched layer exhibited a more pronounced PL increase than a polished electronic-grade diamond sample prepared according to Ref.\cite{Sangtawesin2019a}. This difference indicates that the phenomenon is strongly influenced by the specific surface condition of the diamond.

We also note that similar confocal measurements of the same sample under ambient conditions did not exhibit this increased surface background PL, despite the expectation that the diamond surface is covered by atmospheric adsorbates. This observation indicates that a vacuum environment is necessary for this surface adsorbate-related background PL to manifest.

A recent study reported laser-induced contamination and fluorescence at the diamond–vacuum interface.\cite{parthasarathy2024role} In contrast, the present experiments were performed at substantially lower base pressures, and the morphology of the induced fluorescence differed from that previously reported. These discrepancies indicate that the phenomena observed here likely arise from a different physical origin. Notably, the fluorescence is only observed when the diamond surface is covered with adsorbates and the laser is parked on an NV center, indicating that it likely originates from surface charging effects mediated by interactions with these adsorbates.

Taken together, these results demonstrate that the surface background PL is highly sensitive to surface exposure history and can be reversibly modified through adsorption and desorption processes. These findings underscore the importance of a well-controlled UHV environment for systematic studies of diamond surface effects and their influence on shallow NV center optical properties.

\subsection{\label{sec:UHV_NV_T1_T2}Spin characterizations of single NV centers}

\begin{figure}
    \centering
    \includegraphics{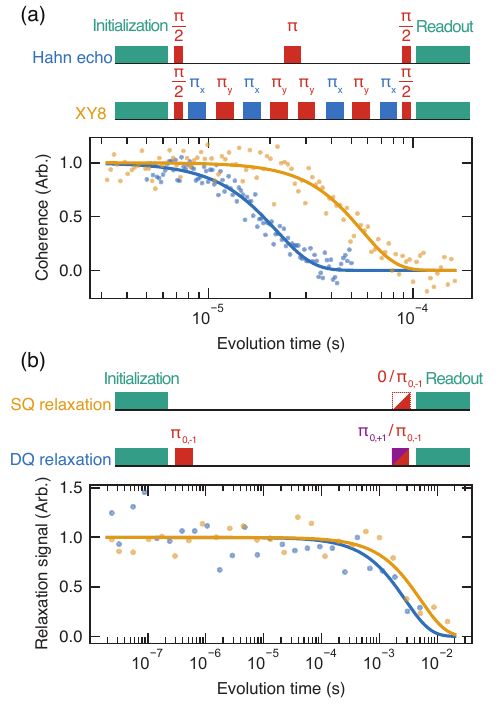}
    \caption{(a) Spin coherence measurements of a shallow NV center in UHV. The Hahn echo measurement (blue) yields $T_2$=21.52$\pm$0.47~$\mathrm{\mu}$s, while the XY8 dynamical decoupling sequence (orange) extends the coherence time to $T_2$=56.33$\pm$1.64~$\mathrm{\mu}$s. The applied magnetic field is 420~G. (b) Longitudinal relaxation measurements of the same NV center. The single-quantum (SQ) $T_1$ (orange) is 5.16$\pm$1.00~ms, and the double-quantum (DQ) $T_1$ (blue) is 2.81$\pm$0.75~ms, measured at a magnetic field of 18~G. The corresponding pulse sequences are shown above each plot. Experimental data (points) are fitted to exponential functions (solid lines).}
    \label{fig:T1_T2}
\end{figure}
% NV #5, depth (17.7, 0.49)

To evaluate the performance of the cryogenic confocal microscope chamber, representative spin relaxation and coherence measurements were performed on shallow single NV centers. 

Spin coherence times were characterized using Hahn echo and XY8 dynamical decoupling sequences (Fig.~\ref{fig:T1_T2}(a)). The Hahn echo measurement provides an estimate of the intrinsic phase coherence time  under refocusing of quasi-static noise with a central \(\pi\)-pulse, while the XY8 sequence extends coherence by suppressing higher-frequency noise components through a train of decoupling \(\pi\)-pulses. Figure~\ref{fig:T1_T2}(a) shows representative results for an NV center with a depth of \(17.7 \pm 0.5\)~nm, where the XY8 sequence increases the NV \(T_2\) by a factor of 2.6 compared to Hahn echo. Dynamical decoupling sequences with more $\pi$-pulses can also be performed to further extend coherence and characterize the spectrum of the noise experienced by the NV centers.\cite{Romach2015}

Longitudinal spin relaxation times were measured in both the single-quantum (SQ) and double-quantum (DQ) bases (Fig.~\ref{fig:T1_T2}(b)). In SQ \(T_1\) measurements, the NV spin is initialized in the \(m_s = 0\) state and, after a variable relaxation time, the population difference between the \(m_s = 0\) and \(m_s = -1\) states is read out using a \(\pi_{0,-1}\) pulse. These measurements probe magnetic noise transverse to the NV axis at the electron spin resonance frequency of the \(m_s = 0 \leftrightarrow -1\) transition. In DQ \(T_1\) measurements, the NV is initialized in the \(m_s = -1\) state, and after relaxation, the population difference between the \(m_s = -1\) and \(m_s = +1\) states is measured by selectively applying either a \(\pi_{0,-1}\) or \(\pi_{0,+1}\) pulse. The DQ basis is sensitive to electric field fluctuations at the \(m_s = -1 \leftrightarrow +1\) transition frequency.\cite{myers2017double} Figure~\ref{fig:T1_T2}(b) shows the SQ and DQ \(T_1\) data for the same NV center as in Fig.~\ref{fig:T1_T2}(a), yielding \(T_{1,\mathrm{SQ}} = 5.16 \pm 1.00~\mathrm{ms}\) and \(T_{1,\mathrm{DQ}} = 2.81 \pm 0.75~\mathrm{ms}\). The ability to access both SQ and DQ relaxation channels provides additional information about the spectral characteristics and possible origins of the local noise environment experienced by shallow NV centers.

These measurements establish that the integrated UHV platform supports quantitative NV spin characterization comparable to conventional confocal systems,\cite{Sangtawesin2019a} while uniquely preserving a well-defined surface environment. 
In future experiments, direct comparison of NV relaxation and coherence times under different surface conditions---such as after thermal annealing or atomic termination via the thermal gas cracker---will allow isolation of noise contributions associated with surface chemistry and reconstruction. This approach will help identify potential noise sources for shallow NV centers, including fluctuating paramagnetic surface states, adsorbate-induced magnetic moments, and surface charge traps. 

\subsection{\label{sec:oxygen_desorption}In situ monitoring of oxygen desorption by XPS}

\begin{figure}
    \centering
    \includegraphics{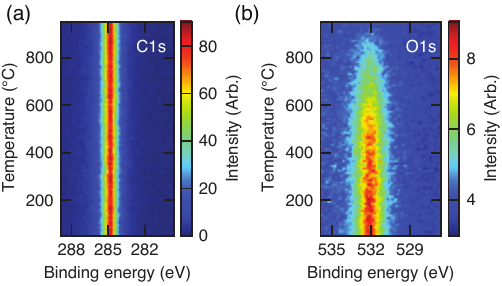}
    \caption{XPS scans of the diamond surface during thermal annealing around the C1s peak (a) and around the O1s peak (b). Sample is heated to above $900\,^\circ\text{C}$ at a rate of around $6\,^\circ\text{C/min}$. The XPS analyzer runs in the ``Peak Monitoring'' mode to allow time-resolved measurements of multiple peaks. The analyzer dwell time at each point is 0.1~s and the pass energy is 100~eV. }
    \label{fig:oxygen_desorption}
\end{figure}

The integration of surface processing and characterization tools within the surface science chamber enables real-time monitoring of diamond surface evolution during thermal treatment. As a representative example, we investigated oxygen desorption from the diamond surface using in situ XPS during high-temperature annealing.\cite{mccloskey2024methods}

In this experiment, the diamond sample was heated to temperatures exceeding $900\,^\circ\text{C}$ while repeatedly acquiring XPS spectra in the vicinity of the C1s and O1s core-level peaks (Fig.~\ref{fig:oxygen_desorption}). Continuous spectral acquisition during annealing allowed the evolution of surface chemical composition to be tracked without interrupting the thermal process. As the sample temperature increased, a systematic decrease in the O1s peak intensity was observed (Fig.~\ref{fig:oxygen_desorption}(b)), indicating progressive desorption of oxygen-containing surface species.

Real-time measurements of multiple core-level peaks are enabled by the ``Peak Monitoring'' mode of the XPS analyzer, in which the analyzer rapidly switches between different energy windows within a single acquisition. During the high-temperature anneal, small drifts in peak positions ($< 3\,\mathrm{eV}$) were detected, likely arising from temperature-dependent charging or work-function shifts. To account for these effects, the C1s peak---associated with the diamond lattice carbon and expected to remain chemically stable during oxygen desorption---was used as an internal reference for normalization. This normalization procedure enables reliable comparison of O1s signal intensity throughout the annealing process.

These measurements demonstrate the capability of the cluster tool to directly correlate thermal processing with changes in surface chemistry under UHV conditions. Real-time observation of oxygen desorption provides insight into the thermal stability of surface terminations and supports reproducible preparation of well-defined diamond surfaces. Such control is essential for systematic studies of shallow NV centers, where surface oxygen termination and residual adsorbates can strongly influence the noise environment and NV charge-state stability.

\section{\label{conclusion}Conclusion}

We have developed and implemented a modular UHV cluster tool that integrates diamond surface preparation and characterization with cryogenic confocal measurements of shallow NV centers. The system combines a load-lock chamber, a dedicated surface science chamber equipped with thermal annealing, a thermal gas cracker, XPS, and LEED, and a cryogenic confocal microscope chamber supporting high–NA optical access, microwave delivery, and bias magnetic field control. A vacuum-transfer architecture enables samples to be processed, characterized, and measured without exposure to ambient conditions, preserving well-defined surface terminations and minimizing uncontrolled contamination.

Representative measurements demonstrate the sensitivity of diamond surface PL to surface preparation and environmental exposure, underscoring the importance of in situ UHV surface control. Spin coherence and relaxation measurements further establish the capability of the platform to quantify how controlled surface treatments influence the local noise environment experienced by shallow NV centers. Real-time XPS measurements directly track oxygen desorption during high-temperature annealing, providing a clear example of precise and reproducible diamond surface engineering within this UHV cluster tool. The ability to reversibly modify surface conditions and immediately probe their impact on NV optical behavior highlights the utility of the integrated platform.

This instrument provides a versatile framework for systematically investigating surface-induced noise, charge stability, and decoherence mechanisms in shallow NV centers. More broadly, it establishes a pathway toward reproducible surface engineering strategies for diamond-based quantum sensing and quantum information applications under well-defined UHV and cryogenic environments.

%%%%%%%%%%%%%%%%%%%%%%%%%%%%%%%%%%%%%%%%%%%%%%%%%%%

\begin{acknowledgments}

We acknowledge useful discussions with Manik Goyal. We thank Mattias Fitzpatrick for his help with the microwave PCB design, Kai-Hung Cheng for his assistance with the bakeout setup, and Berthold Jack and Ali Yazdani for discussions about UHV cluster tool design. This work was supported by the U.S. Department of Energy, Office of Science, Office of Basic Energy Sciences, under Award Number DE-SC0018978, the NSF under the CAREER program (Grant No. DMR- 1752047), and the DARPA DRINQS program (Agreement No. D18AC00015). We also acknowledge the support from Princeton Imaging and Analysis Center.

\end{acknowledgments}

\section*{AUTHOR DECLARATIONS}

\subsection*{Conflict of Interest}
The authors have no conflicts to disclose.

\subsection*{Author Contributions}
\textbf{Zhiyang Yuan:} Conceptualization (equal); Data curation (equal); Formal analysis (equal); Investigation (equal); Methodology (equal); Resources (equal); Software (equal); Validation (equal); Visualization (equal); Writing – original draft (lead); Writing – review \& editing (equal).
\textbf{Sorawis Sangtawesin:} Conceptualization (equal); Data curation (equal); Formal analysis (supporting); Investigation (supporting); Methodology (equal); Resources (equal); Software (lead); Validation (supporting); Visualization (equal); Writing – original draft (supporting); Writing – review \& editing (supporting).
\textbf{Lila V. H. Rodgers:} Data curation (equal); Formal analysis (supporting); Investigation (supporting); Methodology (supporting); Validation (supporting); Visualization (supporting); Writing – review \& editing (supporting).
\textbf{Kalliope Zervas:} Data curation (supporting); Formal analysis (supporting); Investigation (supporting); Validation (supporting); Visualization (supporting); Writing – review \& editing (supporting).
\textbf{James J. Allred:} Conceptualization (supporting); Methodology (equal); Resources (equal); Writing – review \& editing (supporting).
\textbf{Jared Rovny:} Data curation (supporting); Investigation (supporting); Methodology (equal); Writing – review \& editing (supporting).
\textbf{Patryk Gumann:} Conceptualization (equal); Investigation (equal); Methodology (equal); Resources (equal); Writing – review \& editing (supporting).
\textbf{Nathalie P. de Leon:} Conceptualization (lead); Data curation (equal); Formal analysis (equal); Funding acquisition (lead); Investigation (equal); Methodology (equal); Project administration (lead); Resources (lead); Software (equal); Supervision (lead); Validation (equal); Visualization (equal); Writing – original draft (equal); Writing – review \& editing (equal).

\section*{Data Availability Statement}

The data that support the findings of this study are available from the corresponding author upon reasonable request.

\appendix

\section{\label{vacuum_details}Vacuum details}
 The load-lock chamber is evacuated using a turbomolecular pump backed by a dry scroll pump to avoid contamination associated with oil-sealed pumps. The chamber volume is minimized to enable rapid pump-down during sample loading. The load lock is isolated from the surface science chamber by gate valves, allowing samples to be transferred without compromising UHV conditions. During routine operation, the load-lock pressure stabilizes at approximately $5\times10^{-8}$~mbar. The samples are only transferred from the load-lock chamber to the other two chambers after the load-lock pressure reaches the base pressure to minimize contamination in the other two chambers. The surface science and cryogenic confocal microscope chambers operate in the low UHV regime, with base pressures of approximately $5\times10^{-10}$~mbar. Each chamber is maintained by an independent ion pump, with nominal pumping speeds of 300~L/s (surface science chamber) and 100~L/s (cryogenic confocal microscope chamber), respectively. In addition, the surface science chamber is equipped with a non-evaporable getter (NEG) pump to further enhance vacuum performance. 
 
 Achieving the target base pressure requires baking the UHV system at temperatures above $120\,^\circ\text{C}$ for approximately one week to remove adsorbed gases---primarily water---from the chamber surfaces. To facilitate this process, a custom bakeout tent (Hemi Heating) was constructed to enclose the entire system. The tent circulates heated air to ensure a homogeneous temperature distribution across all components while maintaining flexibility for future modifications to the chamber configuration. Components that are sensitive to thermal gradients, such as viewports and electrical feedthroughs, are wrapped in aluminum foil to minimize temperature differentials. Thermocouples are attached to critical locations to continuously monitor temperature throughout the bake. The temperature ramp rate is limited to below $3\,^\circ\text{C/min}$ to prevent thermal stress and potential leak to the system. Following the bakeout, the installed instruments are powered on for degassing to release any residual gases trapped within internal components, including the ion gauges, XPS, LEED, and the gas cracker.

 RGA measurements acquired during chamber bakeout indicate that the chamber base pressure is dominated by hydrogen, likely originating from outgassing of the chamber walls. Hydrogen is not expected to significantly impact shallow NV centers in diamond. The partial pressure of water is more than two orders of magnitude lower than the total chamber pressure, while hydrocarbon species are more than four orders of magnitude lower. Given a base pressure of $5\times10^{-10}$~mbar in the cryogenic confocal microscope chamber, these conditions imply an extremely low flux of contaminant species impinging on the diamond surface. Based on these partial pressures, we estimate that the diamond surface can remain effectively clean for longer than one month under UHV conditions. This estimate is further supported by optical measurements: continuous monitoring of the diamond surface PL in confocal scans shows that an increase in surface PL around the laser spot---used here as an indicator of adsorbate accumulation (see Fig.~\ref{fig:parking_PL})---does not appear for over one month following annealing.

% \nocite{*}
\bibliography{UHV_instrument}% Produces the bibliography via BibTeX.

\end{document}